\documentclass[noshowpacs,preprintnumbers,amsmath,amssymb]{revtex4}
\usepackage{amssymb}
\usepackage{amsmath}
\usepackage{mathrsfs}
\usepackage{graphics}
\usepackage{epsfig}
\usepackage{graphicx}
\usepackage{subfigure}
\usepackage{bm}
\usepackage{nicefrac}
\usepackage{feynmp}

\def\beq{\begin{equation}}
\def\eeq{\end{equation}}
\def\baq{\begin{eqnarray}}
\def\eaq{\end{eqnarray}}

\def\be{\begin{equation}}
\def\ee{\end{equation}}
\def\bea{\begin{eqnarray}}
\def\eea{\end{eqnarray}}

\def\ds{\delta \sigma}

\def\fnl{f_{\rm NL}}
\def\bfnl{\bar{f}_{\rm NL}}

\def\gnl{g_{\rm NL}}
\def\bgnl{\bar{g}_{\rm NL}}

\def\taunl{\tau_{\rm NL}}
\def\btaunl{\bar{\tau}_{\rm NL}}

\def\bhnl{\bar{h}_{\rm NL}}

\def\q{{\bf q}}
\def\x{{\bf x}}
\def\k{{\bf k}}
\def\d{{\rm d}}

\begin{document}
\unitlength = 1mm

\hfill{CERN-PH-TH/2012-193\qquad NORDITA-2012-54}
%\title{Loop corrections to $n$-point functions in single source inflation}
%\title{Loop corrections to the primordial density perturbation and tests of inflation}
\title{Loop corrections and a new test of inflation}
\author{Gianmassimo Tasinato$^{(1)}$, Christian T. Byrnes$^{(2)}$, Sami Nurmi$^{(3)}$, David Wands$^{(1)}$}
% \altaffiliation[Also at ]{Physics Department, XYZ University.}
\affiliation{%
%~ \\
$(1)$ Institute of Cosmology $\&$ Gravitation, University of Portsmouth,
 Portsmouth, PO1 3FX, United Kingdom\\
$(2)$ CERN, PH-TH Division, CH-1211, Gen\`eve 23, Switzerland\\
$(3)$ NORDITA, Roslagstullsbacken 23, SE-106 91, Stockholm, Sweden}%

\begin{abstract}
\noindent
Inflation is the leading paradigm for explaining the origin of primordial density perturbations.
% and the observed temperature fluctuations of the cosmic microwave background.
  However many open questions remain, in particular whether one or more scalar fields were present during inflation and how they contributed to the primordial density perturbation. We propose a new  observational test of whether multiple fields, or only one (not necessarily the inflaton) generated the perturbations. We show that our test, relating the bispectrum and trispectrum, is protected against loop corrections at all orders, unlike previous relations.
\end{abstract}
\maketitle

\smallskip

\noindent
{\bf{Introduction}}:
The statistical distribution of the primordial density field provides a unique opportunity to test our understanding of the origin of the observed Universe. Inflation is the leading paradigm, but many questions about its details remain and it is so flexible that it is unclear if it can ever be ruled out. However it is at least possible to test, in a model independent way, whether a single source present during inflation was responsible for generating density perturbations, or whether multiple sources are required, by exploiting deviations from a Gaussian distribution.
Non-Gaussianity (nonG) parameters are related by distinctive consistency relations, whose structure generally depends on whether perturbations are produced by one or more scalar fields.
These parameters could be measured by forthcoming Planck satellite data
and we may be able to answer fundamental questions about the number of degrees of freedom contributing to the primordial density perturbation.

In single-source scenarios, density perturbations are generated by quantum fluctuations in
a single scalar field, that does {\it not} necessarily correspond to an
inflaton field driving the Hubble expansion. Single-field
inflation is  one particular case of this set-up, in which a single
inflaton field both drives inflation and generates primordial
cosmological perturbations.  More generally, %when discussing single-source models,
we include also set-ups such as curvaton or
modulated reheating scenarios, in the limit that only one
field generates the density perturbation \cite{Suyama:2010uj}.
% {\bf think its enough to cite this one paper, which is the first to use the language "`single source"' and not individually for curvaton etc CB}.
In these models, large nonG of local shape is induced by a single light scalar field, whose
dynamics is important on super Hubble scales.

%becomes important only after inflation ends. In multiple-source scenarios, instead, the density perturbation is produced by
%more than one scalar field. This category includes models of multi-field inflation and multiple-field versions of
 % the aforementioned  non-standard
%other scenarios; all of which may also produce large local nonG.

% DW: Could omit this first consistency relation as it is not relevant to the single-source issue and so obscures our point:
One famous consistency relation associates the squeezed limit of 3-point function with the scale dependence of the 2-point function:
$\fnl = -\nicefrac{5}{12}\,(n_\zeta-1)$. This relation is valid only for single-field (clock) inflation
\cite{Creminelli:2004yq}, and is violated in more general single-source, or multiple-source scenarios.
Another consistency relation, on which we will focus, connects the collapsed limit of 4-point
function to the squeezed limit of the 3-point function:
 \be
  \taunl=\left(\nicefrac{6 \fnl}{5} \right)^2\label{tfe}\,.
  \ee
This equality is
% regarded as being
satisfied at tree-level in single-source scenarios (up to gravitational corrections that can violate it by a small amount \cite{Seery:2008ax},
 see however our Conclusions), but
is generally violated in multiple-source set-ups, leading to the
%more general
Suyama-Yamaguchi inequality $\tau_{\rm NL}\geq \left(\nicefrac{6
\fnl}{5} \right)^2$
\cite{Suyama:2007bg,Smith:2011if,Assassi:2012zq,Kehagias:2012pd}. 
% DW: SAVE SPACE! (see \cite{Assassi:2012zq} for a recent nice paper discussing these topics, and references therein).
Recently it was shown that
the equality (\ref{tfe}) can be broken at an observable level,
{\it even} in single-source scenarios, due to loop corrections \cite{Byrnes:2011ri,Sugiyama:2012tr}.
A popular model which can realise this possibility is the interacting curvaton scenario, several other models also exist \cite{Byrnes:2011ri}.
%A concrete well-known set-up that realizes  this possibility  is the self-interacting curvaton scenario, although other examples exist -- see \cite{Byrnes:2011ri}.

Inflationary observables associated with any given $n$-point ($n$-pt)
function receive loop contributions, in terms of integrals over internal soft momenta that induce logarithmic corrections proportional
to parameters related with higher $n$-pt functions. These
contributions are clearly seen in a
diagrammatic representation of $n$-pt functions in terms of
Feynman-type diagrams \cite{Byrnes:2007tm}. Loop corrections to correlation
functions contribute  to $\fnl$ and $\taunl$, and in single-source scenarios, these corrections
 can combine in such a way to break the equality (\ref{tfe}). As we will discuss,
 the violation of   (\ref{tfe}) can be interpreted as due to the fact that the equality written in this way
  does not  include contributions of soft momentum lines, connecting different $n$-pt functions entering the definitions
  of  %the parameters 
  $\taunl$ and $\fnl^2$. These soft momentum lines are allowed by momentum conservation,
and lead to radiative corrections comparable to the conventional
loop corrections.
% in the limits of collapsed and squeezed configurations.
Accounting for both loop corrections and soft modes,
we will build a new combination of bispectrum,
trispectrum and power spectrum, that leads to an equality satisfied
to all orders in radiative corrections in generic single-source
scenarios. The equality reduces to equation (\ref{tfe}) at
tree-level, and is in general broken in  multiple-field scenarios. Our
result therefore generalizes the Suyama-Yamaguchi relation to all
orders in radiative corrections. Moreover, in the second part, %of this work,
  we will also discuss how
 the soft modes we consider are the source of the inhomogeneity of nonG observables discussed in \cite{Byrnes:2011ri},
 leading to further observational implications for our findings. Finally, in the conclusions we will point out that our generalized
 inequality is preserved by gravitational corrections that spoil (\ref{tfe}).

The role of soft momenta is reminiscent of what happens in
QED, in which a careful inclusion of contributions of both real and
virtual soft photons is crucial for canceling IR divergences in
physical processes  \cite{Bloch:1937pw}. The conceptual idea we develop  here is  similar
to what happens in that context, although the technical implementation will be different.

\smallskip

\noindent
{\bf{Radiative corrections to $n$-pt functions.}} From now on we focus on
a local Ansatz for the primordial curvature perturbation \cite{Wands:2010af}
%\beq
%\label{ans}
%\zeta(\x) = \zeta_G(\x)+\frac{3 \bfnl}{5} \left( \zeta_G^2(\x) -
%\langle  \zeta_G^2(\x) \rangle \right)+\frac{9 \bgnl}{25}
% \zeta_G^{3}(\x)+\frac{81 \bhnl}{125} \left( \zeta_G^4(\x) -6  \langle  \zeta_G^2(\x) \rangle^2 \right)+\ldots\ .
%\eeq
\beq
\label{ans}
\zeta(\x) = \zeta_G(\x)+\frac{3 \bfnl}{5} \zeta_G^2(\x)+\frac{9
\bgnl}{25}
 \zeta_G^{3}(\x)+\frac{81 \bhnl}{125} \zeta_G^4(\x)+\cdots - \langle  \zeta(\x) \rangle\
 ,
\eeq
where $ \zeta_G$ is a Gaussian random fluctuation, with vanishing
ensemble average $\langle\zeta_G\rangle\,=\,0$. The local Ansatz
assumes the parameters $\bfnl$, $\bgnl$ and $\bhnl$ to be constant. For our arguments, 
we will assume that these parameters  are sufficiently large to be observable: in this case,  slow-roll suppressed contributions to non-Gaussianity,  associated with the intrinsic non-Gaussianity of the fields under consideration, provide only small corrections to our results  in the relevant momentum limits,  and can be neglected.  
%We use a bar to emphasize that these parameters, as defined by
%(\ref{ans}), characterize tree-level amplitudes of 3-, 4-,
%and 5-pt functions respectively.
% {\bf GT we've to decide how to call the parameters entering in the previous formula.
%    David suggests to call them $c_2$, $c_3$ etc. The negative side of this is
%  that   it would seem
%  very strange for the reader to meet these parameters, instead of the usual
%  $\fnl$ etc. The positive side is  that doing this way  would certainly catch the reader attention on this issue. We've to think about}
%  % Ergodic theorem ensures that spatial averages coincide with ensemble averages, if
%the space is sufficiently large.
% The parameters $\fnl$, $\gnl$ are assumed
%to be constant.% and as defined above are only correct at tree level.
We focus on single-source scenarios, in which  the curvature
perturbation is generated by  a single scalar $\sigma$ and with observably large nonG. It is then possible to rewrite the local Ansatz
(\ref{ans}) in terms of an expansion of a suitable classical
function $N(\sigma)$,
  \bea
  \label{expan1}
  \zeta(\x) &=& N(\sigma_0+\delta \sigma(\x))-\langle N(\sigma_0+\delta \sigma(\x))  \rangle
  %\\
 % &=&N'(\sigma_0) \delta \sigma(\x) +\frac12 N''(\sigma_0) \left(
  %\delta \sigma^2(\x) -\langle\delta \sigma^2(\x)\rangle \right) %\nonumber\\&&
  %+ \frac16 N'''(\sigma_0) \delta \sigma^3(\x)+\dots \label{expan2}\ .
  %\\
  \,=\,  \sum_{n=1}^{\infty}\frac{N^{(n)}(\sigma_0)}{n!}\,\left(\delta\sigma(\x)^n - \langle \delta\sigma(\x)^n\rangle\right)
 \eea
This expansion is similar to the $\delta N$ approach  of
\cite{Lyth:2005fi} but for our purposes it is not necessary to
specify $N(\sigma)$ further. $\sigma_0$ denotes a homogeneous, time
dependent background solution
% DW: SAVE SPACE , as dictated by the symmetries of FRW space,
and the fluctuations $\delta \sigma(\x)$ are Gaussian with
zero mean. Making the identification $N'(\sigma_0) \delta
\sigma(\x)\,=\,\zeta_G(\x)$, a comparison of the equations
(\ref{ans}) and (\ref{expan1}) yields the relations $\nicefrac{6
\bfnl}{5}\,=\,\nicefrac{N''}{N'^2}$, $\nicefrac{54
\bgnl}{25}\,=\,\nicefrac{N'''}{N'^3} $, etc \cite{Lyth:2005fi}.
%{\bf DW: shouldn't we cite Lyth+Rodriguez here?}

The general definitions for the parameters $\fnl$ and $\taunl$ at
tree-level and beyond are given in appropriate squeezed and
collapsed limits by
 \baq
 {f}_{{\rm NL}} &\equiv&\frac{5}{12}\,\lim_{k_1\to0}\frac{B(k_1,k_2,k_3)}{P(k_1)
  P(k_2)}%=\frac{5}{6}\,\frac{N''}{N'{}^2}
  \ ,
  \label{pfnlt}\\
  \label{ptnlt}  {\tau}_{{\rm
  NL}}&\equiv&\frac{1}{4}\,\lim_{{k_{12}}\to0} \frac{T(k_1,k_2,k_3,k_4,k_{12},k_{13})}{P(k_{12}) P(k_1)
  P(k_3)}%=\frac{N''^2}{N'{}^4}\nonumber
  \ ,
  \eaq
where
%\bea
$\langle \zeta_{{\bf k}_1}\zeta_{{\bf k}_2}
%\zeta_{{\bf k}_3}
\rangle = (2\pi)^3 P
 \delta({\bf k}_1+{\bf
k}_2)
 $,
%\\
$\langle \zeta_{{\bf k}_1}\zeta_{{\bf k}_2}\zeta_{{\bf
k}_3}\rangle\,=\,(2\pi)^3 B
 \delta({\bf k}_1+{\bf
k}_2+{\bf k}_3)
$,  $ \langle \zeta_{{\bf k}_1}\zeta_{{\bf
k}_2}\zeta_{{\bf k}_3}\zeta_{{\bf k}_4}\rangle \,=\,(2\pi)^3
T
%(k_1,k_2,k_3,k_4,k_{12},k_{13})
\delta({\bf k}_1+{\bf k}_2+{\bf
k}_3+{\bf k}_4)
$,
and $k_{ij}=|{\bf k}_i+{\bf k}_j|$. Notice that at tree-level $\fnl$
defined by equation (\ref{pfnlt}) reduces to the parameter $\bfnl$
entering the local Ansatz (\ref{ans}). At tree level one has
$\bar{f}_{{\rm NL}} \,=\, \nicefrac{5\,N''}{6\,N'{}^2}$, $
\bar{\tau}_{{\rm NL}}\,=\,\nicefrac{N''^2}{N'{}^4}$, from which
equality (\ref{tfe}) follows immediately. Let us stress that, in this work, we focus on the case in which
 the tree-level bar quantities are constant and do not depend on momenta. 
However, when 1-loop
contributions are added to the tree-level results, one finds
%~\footnote{Notice that loop
%contributions induces scale dependence to local nonG parameters. If
%$h_{\rm{NL}}$ is large, $n_{\fnl}$ and  $n_{\gnl}$ can be large
%enough to be observable.}
   \bea
  \label{fnl_loop}
  \fnl^{\rm loop}&=&\bfnl
  -\frac{18}{25}\left(2 \bfnl^3-3 \bfnl \bgnl-3 \bhnl\right)\,\bar{{\cal P}} \ln{\left(\frac{k}{k_{IR}}\right)}\ ,\\
  \taunl^{\rm loop}&=&\bar{\tau}_{\rm NL}-\frac{324}{625}
  \left(8 \bfnl^4-12 \bfnl^2 \gnl-9 \bgnl^2-12 \bhnl \bfnl\right)\,\bar{{\cal P}} \ln{\left(\frac{k}{k_{IR} }\right)},
  \nonumber
  \eea
%[Please double check this, im not sure about numerical factors but there must be a $h_{NL}$ term, in end i still get (\ref{loop-violation}), i instead find -CB] {\bf GT it's true that in my formulae the $\hnl$ was missing: i removed it because it didn't fit in two columns. Now that we're back in one column, i re-put it. The above is my version of the result: I find results slightly different from yours, i think there was a typo in yours. Indeed take care that your expression for $\taunl$ loop, last term, can't be correct because can't depend on just $\fnl^2$.}
 % \bea
  %\fnl^{loop}&=&\fnl
 % +\frac{18}{25}\left(3h_{NL}+3 \fnl \gnl-2 \fnl^3\right)\,{\cal P}_\zeta \ln{\frac{k}{k_{IR}}}\\
 % \taunl^{loop}&=&\taunl+\frac{324}{625}\left(12 h_{NL}\fnl +12\gnl\fnl^2+9\gnl^2-8\fnl^2\right)\,{\cal P}_\zeta \ln{\frac{k}{k_{IR}}}\nonumber
 % \eea
%
where $2\pi^2{\cal
P}=k^3P(k)$, and we neglect its weak scale dependence. Hence
 \be\label{loop-violation}
 \taunl^{\rm loop}\,=\,\left(\frac{6 \fnl^{\rm loop}}{5}\right)^2\,
 \left[1+\frac{81\,\bgnl^{\,2}}{25\bfnl^{\,2}\rule{0pt}{2.2ex}}\,\bar{\cal P} \ln{\left(\frac{k}{k_{IR}}\right)} \right]
 \ee
showing that the consistency relation (\ref{tfe}) is violated
already at one loop in single-source models, if the tree-level
$\bar{g}_{\rm NL}$ is non-vanishing.
More precisely, the consistency relation holds only on the scale $k_{\rm IR}$ at which the tree-level quantities are defined and the loops are absent. Moving away from this scale, the loop corrections become non-zero leading to a violation of the consistency relation.
 If the nonG parameters
are large, this violation of the consistency relation can be
observed by the Planck satellite. 
%{\bf GT added  next part} 
 As a representative example, assume $\bar{f}_{\rm NL}\,=\,20$, and $\bar{g}_{\rm NL}\,=\,8\,\cdot\,10^5$, close to the 
 upper bound set by WMAP. See e.g. \cite{Enqvist:2008gk}
   for explicit  examples of   models that theoretically under control, and that can lead to such
 a large hierarchy between tree level values of $\bar{f}_{\rm NL}$ and  $\bar{g}_{\rm NL}$.
   Without  including loop corrections (the square parenthesis in (\ref{loop-violation}))
  one would find $\tau_{\rm NL}\,=\,576$:
 too small value to be observed in the near future, since
   the forecasted Planck  constraint is   $\tau_{\rm NL} \gtrsim 1500$   at 1-sigma error bar, in absence of detection \cite{Smidt:2010ra}
   (see also \cite{Kogo:2006kh} for an analysis suggesting that CMB data  might lead  to even lower  values for the detectability of $\taunl$). Including loops, instead,
   the value of  $\tau_{\rm NL}$ becomes large enough to be detectable:    $\tau_{\rm NL}\,\sim\, 3600$ 
   setting  $\bar  {\cal P} \sim 10^{-9}$ to match COBE normalization, and assuming  $ \ln \left(\nicefrac{k}{k_{IR}}\right) $  of order one. Hence in this situation loop corrections can really make the difference and render $\tau_{\rm NL}$ detectable even in the single source case.

%  As a representative example, assume that $ \bar{g}_{\rm NL}  \sim 8 \cdot 10^5$, close to the upper bound set by WMAP. Taking $\bar{f}_{\rm
% NL} \sim 10$, setting
% $\bar  {\cal P} \sim 10^{-9}$ to match COBE normalization, and assuming  $ \ln \left(\nicefrac{k}{k_{IR}}\right) $  of order one, one finds
 %that the parenthesis in the right hand side of (\ref{loop-violation}) can be  of order 100 -- drastically spoiling the tree-level consistency relation. Hence, given that current data do not forbid loop corrections from being large,   
  % it is imperative to theoretically clarify the issue,   to be ready for correctly interpreting the forthcoming results of the Planck satellite. 

  A couple of considerations on to the physical validity of our calculations.  One might be worried with respect to the fact that, choosing such
  a large value for $\bar{g}_{\rm NL}$ as in the example above,
    the loop contribution  to some $n$-point functions dominate over the tree level term: in particular the third order term in (\ref{ans}) can  be larger than the second order term. This raises questions on whether our calculation is under control
    at the technical level, in particular whether a perturbative approach makes sense. Fortunately it does (provided that terms beyond $\bar{g}_{\rm NL}$  in (\ref{ans}) do not grow too rapidly). Indeed, it is possible to show that only one non-linearity parameter can be associated with each external line in a  loop diagram \cite{Byrnes:2007tm}, and so for the power spectrum the largest possible  power of $\bar{g}_{\rm NL}$  is $\bar{g}_{\rm NL}^2$   (in general for an $n$-point function it is $\bar{g}_{\rm NL}^n$). 
  For the example we consider, this implies that the one-loop term always dominates over higher loops: higher loops to the trispectrum contain at most four powers of of $\bar{g}_{\rm NL}$ but they are also suppressed by higher powers of $\bar{\cal P}$ than the one-loop term in such a way that the total value is smaller. To be specific, there is a 3 loop contribution to $\taunl$ which goes like $\bar{g}_{\rm NL}^4 \bar{\cal P}^3$, and so although $ \bar{g}_{\rm NL}^2 \bar{\cal P}\gtrsim 1$ one has $ \bar{g}_{\rm NL}^2 \bar{\cal P}^2\ll1$ and hence the one loop term dominates.

  %   For the example we consider, this implies  that the one-loop term always dominates over higher loops: higher loops contain at most two
    % powers of $\bar{g}_{\rm NL}$; on the other hand, they will be suppressed by higher powers of $\bar{\cal P}$ than the one-loop term.  
    
A second concern is about the fact that, in the particular case of pure single field inflation, it has been shown that % the effects of  
 loop corrections % to observables associated with cosmological perturbations 
   can
be  absorbed   in a redefinition of background  quantities, by a proper choice of physical coordinates: see \cite{Gerstenlauer:2011ti,Giddings:2011zd,Giddings:2010nc} for the first papers   on these topics. More in particular,  the basic idea is the following: since field perturbations are necessarily adiabatic in 
single field inflation, they span the direction of the homogeneous, classical inflationary trajectory. It can be shown that,  by means of a change of coordinates, a suitable shift on this background  trajectory can be made  such 
   to compensate the effects of loop corrections to  observable quantities. This fact
    is however true only for pure single field inflation, and {\it does not apply} in our more general context of single source models that lead
    to large nonG. In this case, indeed, isocurvature fluctuations span directions perpendicular to the homogeneous one, and consequently the corresponding loop effects can not be re-adsorbed by any background redefinition. Hence, the loop effects that we are considering in this paper are fully physical and well defined.

On the other hand,
 while being
 well-defined and consistent, the standard way of
calculating loop corrections {\it individually} to the power
spectrum, bispectrum and trispectrum, and then taking appropriate
ratios to define the non-linearity parameters beyond tree-level, can
miss important physical contributions.
% SN: Modified the rest of the paragraph.
Indeed, these loop corrected quantities correctly characterize the
ratios of individually measured bispectrum, or trispectrum, and
power spectrum. However,
% in terms of soft modes connecting  different $n$-pt functions entering into definitions of nonG parameters.
%Indeed, when considering combinations  of $n$-pt functions
when {\it simultaneously} measuring combinations of $n$-pt
functions, such as the ratios in (\ref{pfnlt}) and (\ref{ptnlt}),
%in
%appropriate collapsed and squeezed limits
%{\bf only then, i would say always, this also means the following sentence may have to be modified CB} {\bf GT this's delicate: i think that soft modes between diagrams play a clean role only when dealing with collapsed limits of n-pt functions. Only then it's clear that they can't be distinguished from the internal soft momentum line. I would like to leave the sentence as it is, in any case it's not wrong },
%as in equations (\ref{pfnlt}) and (\ref{ptnlt}),
one should allow for the inclusion of soft lines connecting distinct
$n$-pt functions.  Although  momentum is of course conserved, no
detector is sensitive enough to probe these soft lines:
% and distinguish them from the
%internal soft lines running within each diagram;
their contribution is physical and must be included.
%{\bf GT added next sentence}
This observation suggests that, besides considering the relation
(\ref{tfe}), by including the effects of soft modes it is possible
to build a new observable combination of $n$-pt functions that leads
to an equality protected against radiative corrections. See also \cite{Giddings:2011zd} for the slow-roll suppressed
       effect of soft modes on the power spectrum in single clock inflation. 
   % We are
   %going to discuss this possibility in the next sections.

  %%%%%

%This method, although perfectly
 %fine since it generalizes to the case of loops the procedure adopted at tree level, nevertheless can miss important contributions,
 %in terms of soft  modes that connect different $n$-pt functions.

\smallskip

\noindent {\bf{Diagrammatic approach to loop corrections.}} It is
illuminating to discuss the role of soft modes diagrammatically, first in a
simple example, and then applying our observations  to equality (\ref{tfe}). We
implement the diagrammatic approach of \cite{Byrnes:2007tm}.
% to which we refer the reader for  more details on the diagrammatic method to loop corrections to inflation.
We use solid dots to mark external momenta ${\bf k}_i$,
with the number of attached propagators to each vertex
(corresponding to $P$) giving the number of derivatives of the
function $N$, defined by equation (\ref{expan1}), associated to the
vertex. There are no internal vertices since we assume that
$\delta\sigma$ is Gaussian. The  numerical factors are the
total for each diagram relative to the tree-level term (which may
have some possible permutations), and are given by the numerical
factor of the given diagram ($\nicefrac12$ if a dressed vertex,
otherwise unity at one loop or tree-level), times the number of
distinct ways in which the loop can be drawn onto the tree-level
diagram.
% DW: COULD OMIT THESE DIAGRAMS FROM A SHORT LETTER VERSION
%For example, up to one loop the bispectrum is given by (the first
%diagram is the unique tree-level diagram)
%%
%\bea\label{B-diag} \parbox{6mm}{
%\begin{fmffile}{B}
%\begin{fmfgraph}(5,3)
%\fmfbottom{i1,i2,i3}
%\fmf{plain}{i1,i2,i3}
%\fmfdot{i1,i2,i3}
%\end{fmfgraph}
%\end{fmffile}}
%+\frac22 \;\;\;\parbox{6mm}{\begin{fmffile}{B-dl}
%\begin{fmfgraph}(5,3)
%\fmfbottom{i1,i2,i3}
%\fmf{plain}{i1,i2,i3}
%\fmf{plain}{i1,i1}
%\fmfdot{i1,i2,i3}
%\end{fmfgraph}
%\end{fmffile}}
%+\frac12\; \parbox{6mm}{
%\begin{fmffile}{B-dc}
%\begin{fmfgraph}(5,3)
%\fmfbottom{i1,i2,i3}
%\fmf{plain}{i1,i2,i3}
%\fmf{plain}{i2,i2}
%\fmfdot{i1,i2,i3}
%\end{fmfgraph}
%\end{fmffile}}
%+2\;\; \parbox{6mm}{
%\begin{fmffile}{B-side}
%\begin{fmfgraph}(5,3)
%\fmfbottom{i1,i2,i3}
%\fmf{plain}{i1,i2,i3}
%\fmf{plain,left}{i1,i2}
%\fmfdot{i1,i2,i3}
%\end{fmfgraph}
%\end{fmffile}} +1\; \parbox{6mm}{
%\begin{fmffile}{B-long}
%\begin{fmfgraph}(5,3)
%\fmfbottom{i1,i2,i3}
%\fmf{plain}{i1,i2,i3}
%\fmf{plain,left}{i1,i3}
%\fmfdot{i1,i2,i3}
%\end{fmfgraph}
%\end{fmffile}}
%\eea

\begin{figure}[h!]
  \begin{center}
    \includegraphics[width=12 cm, height= 1.6 cm]{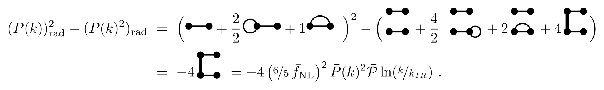}
  \end{center}
  \vspace{-0.6 cm}
  \caption{The difference between observables $( P(k) )_{\rm rad}^2$ and $( P(k)^2)_{\rm rad}$}
  \label{fig:1}
\end{figure}

Let us illustrate the role of soft modes, by considering radiative
corrections to the square of the power spectrum as an example. We
denote with  $( \dots )_{\rm rad}$ the sum of tree-level and
radiative contributions to a given quantity. Figure \ref{fig:1}
depicts diagrammatically the
%We estimate, to first order in radiative
%corrections, the
difference between $(P)_{\rm rad}^2$, associated
with  the observable $\langle\zeta(\k)\zeta(\k')\rangle^2$, and
$(P^2)_{\rm rad}$, denoting another observable,
$\langle\zeta^2(\k)\zeta^2(\k')\rangle$.
%Diagrammatically:
The final diagram represents a 4-pt function with a soft internal
line (drawn thicker) connecting two $N''$ vertices (which can be
done in four ways). This contribution cannot be %observationally
distinguished from the product of two disconnected 2-pt functions
and must be included in $(P^2)_{\rm rad}$. So,
 accounting for loop corrections only,  would not give the correct
result for the observable $(P^2)_{\rm rad}$
%(\ref{Pvariance})
, showing the
importance of the soft modes.

%\smallskip

Contributions to radiative  corrections associated with soft lines
connecting different diagrams play a crucial role for characterizing
equality (\ref{tfe}) beyond tree-level. In order to analyze it diagrammatically, it is
convenient to re-express it in terms of  square of the bispectrum, and
the product of the trispectrum with the power spectrum. At tree-level,
(\ref{tfe}) reads $  \bar{B}^2_{\,\,k_1\rightarrow0}\, = \, \bar{T} \bar{P}(k_5)_{\,\,k_{12}\rightarrow0,\,k_5\rightarrow0}$, 
where here and in what follows, to avoid ambiguities,  we assume that, in the zero momentum limit, all soft momenta coincide: $k_1=k_{12}=k_5\to0$.
% {\bf Perhaps we could remove this and next equation to save space? the tree level equality is most easily derived without diagrams anyway CB} Then, at tree-level,
%(\ref{tfe}) diagrammatically reads
%\bea  \bar{B}^2_{\,\,k_1\rightarrow0} =\;\;\;\; \; \parbox{10mm}
%{\begin{fmffile}{Bsquared}
%\begin{fmfgraph*}(5,10)
%\fmfleft{i1,i2,i3}
%\fmfright{o1,o2,o3}
%\fmf{plain}{i1,i2,i3}
%\fmf{plain}{o1,o2,o3}
%\fmflabel{$k_3$}{i1}
%\fmflabel{$k_2$}{i2}
%\fmflabel{$k_1$}{i3}
%\fmflabel{$k_3$}{o1}
%\fmflabel{$k_2$}{o2}
%\fmflabel{$k_1$}{o3}
%\fmfdot{i1}
%\fmfdot{i2}
%\fmfdot{i3}
%\fmfdot{o1}
%\fmfdot{o2}
%\fmfdot{o3}
%\end{fmfgraph*}
%\end{fmffile}}  = \bar{T} \bar{P}(k_5)_{\,\,k_{12}\rightarrow0,\,k_5\rightarrow0}
 % =\;\;\;\;\; \parbox{10mm}{
%\begin{fmffile}{PTtree}
%\begin{fmfgraph*}(5,10)
%\fmfleft{i1,i2,i3}
%\fmfright{o1,o2,o3}
%\fmf{plain}{i2,i3,o3,o2}
%\fmf{plain,tension=1}{i1,o1}
%\fmfdot{i1}
%\fmfdot{i2}
%\fmfdot{i3}
%\fmfdot{o1}
%\fmfdot{o2}
%\fmfdot{o3}
%\fmflabel{$k_5$}{i1}
%\fmflabel{$k_1$}{i2}
%\fmflabel{$k_2$}{i3}
%\fmflabel{$k_6$}{o1}
%\fmflabel{$k_4$}{o2}
%\fmflabel{$k_3$}{o3}
%\end{fmfgraph*}
%\end{fmffile}}
 % \,= \, 4 \left(\nicefrac{6}{5}\,\bfnl\right)^2 {\bar P}(k_2)^2 \bar{P}^2_{k_5\rightarrow0}\,. \eea
Including loop corrections to $P$, $B$  and $T$,
%, see Eqs.~(\ref{P-diag}-\ref{T-diag}),
 in appropriate squeezed and collapsed
limits,
 %and combining the
%diagrams together as follows, in appropriate squeezed and collapsed
%limits, we find diagrammatically
 the relation is
\bea (B)_{\rm rad}^2-( P )_{\rm rad} (  T )_{\rm rad}
 %=- 4\; \parbox{6mm}{\begin{fmffile}{T-top}
 %\begin{fmfgraph}(5,5)
 %\fmfleft{i1,i2}
 %\fmfright{o1,o2}
 %\fmf{plain}{i1,i2,o2,o1}
 %\fmf{plain,left}{i2,o2}
 %\fmfdot{i1,i2,o1,o2}
 %\end{fmfgraph}
 %\end{fmffile}} \; \parbox{6mm}{
 %\begin{fmffile}{P}
 %\begin{fmfgraph}(5,3)
 %\fmfbottom{i1,i2}
 %\fmf{plain}{i1,i2}
 %\fmfdot{i1,i2}
 %\end{fmfgraph}
 %\end{fmffile}}
 \,=\,-4\left(\nicefrac{54}{25}\,\bgnl
\right)^2 \bar{P}(k)^2\bar{P}_{q\rightarrow0}^2\bar{{\cal
P}}\ln(\nicefrac{k}{k_{IR}})\ \,. \eea
So, the equality is
broken by a term proportional to tree-level $\bar{g}_{NL}$, as discussed in the
previous section. However, a straightforward calculation
shows that the relation
  \bea (B^2)^{\rm rad}_{k_1\to0}&=&( P T)^{\rm rad}_{k_{12}\to0,\,k_5\to0}
  \nonumber\\
  &=& %4 \left(\nicefrac{6}{5}\bfnl\right)^2\bar{P}_{\rm hard}^2\bar{P}_{\rm soft}^2
  \bar{B}^2_{\,\,k_1\rightarrow0}
  \Big[1 +
  \left(\nicefrac{6}{5}\right)^2\Big(\nicefrac{3\bhnl}{\bfnl}
  % \nonumber \\ &&
  + \nicefrac{9\bgnl^2}{4\bfnl^2}+15\bgnl+6\bfnl^2\Big){\bar {\cal
  P}}\ln(\nicefrac{k}{k_{IR}})\Big]\ , \label{newequa}
  \eea
  instead leads to an equality that
{\it is}  preserved by radiative corrections. 
 This  new equality holds with for models leading to large nonG of local type, as the ones on which we are
  focussing our attention  in this paper. 
  We have neglected all non-local contributions associated to eventual non-Gaussianity present at horizon crossing. Assuming canonical kinetic terms, this is well justified as  the neglected  contributions are slow roll suppressed. 
 In  the first line of eq. (\ref{newequa}), we send to zero a momentum line (denoted with $k_1$) in each  of the bispectra in the left hand side; in the right hand side, we send to zero the internal momentum line of the trispectrum denoted with $k_{12}$, as well as the momentum $k_5$ characterizing the power spectrum. As explained above, all these momenta are made vanishing with the same rate, 
 and coincide in the zero momentum limit.  
When evaluating the previous
quantities,  it is crucial to include diagrams describing
  contributions of soft modes connecting different elements of each combination (drawn thickly in the diagrams below),
  respectively the 3 diagrams for $( B^2 )_{\rm rad}$  and 2 for  $( P T )_{\rm rad}$.
\begin{figure}[h!]
  \begin{center}
    \includegraphics[width=7 cm, height= 1.0 cm]{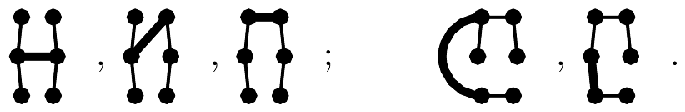}
  \end{center}
  \vspace{-0.6 cm}
  %\caption{The difference between observables $( P(k) )_{\rm rad}^2$ and $( P(k)^2)_{\rm rad}$}
  %\label{fig:1}
\end{figure}
%
%  {\bf GT here we should put the second figure}
%
%\bea\; \parbox{8mm}{\begin{fmffile}{Bsquared-gg}
%\begin{fmfgraph}(4,8)
%\fmfleft{i1,i2,i3}
%\fmfright{o1,o2,o3}
%\fmf{plain}{i1,i2,i3}
%\fmf{plain}{o1,o2,o3}
%\fmf{plain,width=2.3,fore=red}{i2,o2}
%\fmfdot{i1}
%\fmfdot{i2}
%\fmfdot{i3}
%\fmfdot{o1}
%\fmfdot{o2}
%\fmfdot{o3}
%\end{fmfgraph}
%\end{fmffile}},\;\; \parbox{8mm}{
%\begin{fmffile}{Bsquared-gff-cross}
%\begin{fmfgraph}(4,8)
%\fmfleft{i1,i2,i3}
%\fmfright{o1,o2,o3}
%\fmf{plain}{i1,i2,i3}
%\fmf{plain}{o1,o2,o3}
%\fmf{plain,width=2.3,fore=red}{i2,o3}
%\fmfdot{i1}
%\fmfdot{i2}
%\fmfdot{i3}
%\fmfdot{o1}
%\fmfdot{o2}
%\fmfdot{o3}
%\end{fmfgraph}
%\end{fmffile}},\;\; \parbox{8mm}{
%\begin{fmffile}{Bsquared-ffff-cross}
%\begin{fmfgraph}(4,8)
%\fmfleft{i1,i2,i3}
%\fmfright{o1,o2,o3}
%\fmf{plain}{i1,i2,i3}
%\fmf{plain}{o1,o2,o3}
%\fmf{plain,width=2.3,fore=red}{i3,o3}
%\fmfdot{i1}
%\fmfdot{i2}
%\fmfdot{i3}
%\fmfdot{o1}
%\fmfdot{o2}
%\fmfdot{o3}
%\end{fmfgraph}
%\end{fmffile}} ;
%\qquad\qquad
%\parbox{8mm}{
%\begin{fmffile}{PT-gff-c}
%\begin{fmfgraph}(4,8)
%\fmfleft{i1,i2,i3}
%\fmfright{o1,o2,o3}
%\fmf{plain}{i2,i3,o3,o2}
%\fmf{plain,tension=1}{i1,o1}
%\fmf{plain,left=1,width=2.3,fore=red}{i1,i3}
%\fmfdot{i1}
%\fmfdot{i2}
%\fmfdot{i3}
%\fmfdot{o1}
%\fmfdot{o2}
%\fmfdot{o3}
%\end{fmfgraph}
%\end{fmffile}},\; \;\parbox{8mm}{
%\begin{fmffile}{PT-ffff-c}
%\begin{fmfgraph}(4,8)
%\fmfleft{i1,i2,i3}
%\fmfright{o1,o2,o3}
%\fmf{plain}{i2,i3,o3,o2}
%\fmf{plain,tension=1}{i1,o1}
%\fmf{plain,width=2.3,fore=red}{i1,i2}
%\fmfdot{i1}
%\fmfdot{i2}
%\fmfdot{i3}
%\fmfdot{o1}
%\fmfdot{o2}
%\fmfdot{o3}
%\end{fmfgraph}
%\end{fmffile}}.
%  \eea

\noindent
The three first diagrams are 6pt functions consisting of two 3-pt functions connected by a soft line. Analogously to the case of 2-pt function in Fig. \ref{fig:1}, they contribute to $(B^2)_{\rm rad}$ but do not appear in the square of the 3-pt function $(B_{\rm rad})^2$. The soft terms thus generate a non-vanishing variance for the bispectrum, $(B^2)_{\rm rad} - (B_{\rm rad})^2\neq 0$, and similar comments apply for $(PT)_{\rm rad}$ on the right hand side of equation (\ref{newequa}).  Including the soft diagrams with their associated numerical coefficients 
% When including these diagrams  with their associated
%numerical factors 
 \cite{Byrnes:2007tm}, one fulfills the equality
(\ref{newequa}),  $( B^2)_{\rm rad} \,=\,( P T)_{\rm rad}$, that
%should be regarded as the most convenient generalization of
 generalizes the
equality (\ref{tfe}) to first order in radiative corrections. The key  observation is that, in order to define an
equality that remains valid when radiative corrections are included,
one should consider radiative contributions to the entire
combination of $( B^2)_{\rm rad}$ and
 $( P T)_{\rm rad}$, that include both loop corrections and contributions from soft modes connecting different diagrams.
%{\bf GT added next sentence} 
   The new combination (\ref{newequa}) can be considered as a new inflationary observable:
   when going  beyond tree level in a loop expansion,
   it allows
    % besides the relation of eq.
   %(\ref{tfe}),
   % allowing
      to probe  nonG parameters
       in a different way with  respect to relation (\ref{tfe}).
       %  {\bf GT added next paragraph, but  not sure about it, looks a bit useless: Changed to SN's suggestion}
      % The three first diagrams are 6pt functions consisting of two 3pt functions connected by a soft line. Analogously to the case of 2pt function in Fig. 1, they contribute to $(B^2)_{\rm rad}$ but do not appear in the square of the 3pt function $(B_{\rm rad})^2$. The soft terms thus generate a non-vanishing variance for the bispectrum, $(B^2)_{\rm rad} - (B_{\rm rad})^2\neq =0$, and similar comments apply for $(PT)_{\rm rad}$ on the right hand side of equation (\ref{newequa}).  Including the soft diagrams with their associated numerical coefficients allows the equality to be preserved.
%        From a different perspective, our results can be seen as follows: the diagrams  above clearly represent tree level, connected
%        contributions to a 6pt function. Each side of eq. (\ref{newequa}) contains all the possible loop contributions to disconnected parts
%        of a 6pt function (respectively, the 3+3 and 4+2 pieces),  completed by the tree-level diagrams providing the  connected parts to it. 
 %       The connected parts in the figure above must be included to fulfill 
  %      equality  (\ref{newequa}). 
        %       beyond tree level in a loop expansion.

\smallskip

\noindent {\bf{An alternative approach to  radiative corrections. }} We reconsider the problem from another point of view,  which
allows a straightforward generalization of our results to all orders
in radiative corrections,  and emphasizes the connection to
inhomogeneities of non-Gaussian observables \cite{Byrnes:2011ri}.
%First, some preliminary material.
First, consider a wavenumber $Q < a H$ which defines a length scale smaller
than the observed universe. The fluctuations in (\ref{expan1}) can
be divided into long wavelength (LW) and short wavelength (SW)
components with respect to this scale as
 \bea
 \delta \sigma(\x) &= &\int_{q > Q} \frac{\d\q}{(2\pi)^3}\,e^{i \q\cdot \x} \delta \sigma(\q)+\int_{q < Q} \frac{\d
  \q}{(2\pi)^3}\,e^{i\q\cdot\x}\delta \sigma(\q)\nonumber\\
  &\equiv& \ds_s (\x)+\ds_L (\x)\ . \label{split}
  \eea
% DW:
For a Gaussian field,
the LW fluctuations are uncorrelated with the SW modes,
$\langle\delta\sigma_L(\k)\delta\sigma_s(\k')\rangle=0$, and they both have
vanishing ensemble averages,
$\langle\delta\sigma_L\rangle=\langle\delta\sigma_s\rangle=0$. Up to cosmic
variance, the ensemble averages correspond to spatial averages over
the full observable sky.

For measurements probing wavenumbers $k > Q$,  or equivalently
regions of size smaller than  $\nicefrac{1}{Q}$,  the LW modes
$\delta \sigma_L$ act as an approximately homogeneous background
for $\delta \sigma_s$. Indeed, the average of $\delta \sigma_{L}$
computed over a spherical patch of volume $V_{\x_0}$, with the
origin located at a fiducial point $\x_0$, is given by
\cite{Byrnes:2011ri} $  \langle\delta \sigma_L(\x)\rangle_{{\rm x_0}}\,\simeq\, \delta
  \sigma_{L}({\bf x}_0)$.
The average $ \langle\delta \sigma_L(\x)\rangle_{{\x_0}}$
consequently depends on the location $\x_0$ of the patch. In
general, the contribution of long-wavelength modes differs from
patch to patch, generating variations in observables  evaluated in
different subhorizon patches across the sky. This has a cumulative
effect, leading to  a log-enhanced variance of the long-wavelength
fluctuations over the entire observable universe (${\cal P}_{\sigma}$ denotes the spectrum of the
fluctuations $\delta\sigma$):
  \beq\label{loge}
  \langle\delta \sigma_L^2(\x)\rangle = \int\limits_{k_{IR}< q < Q}\frac{\d\q}{(2\pi)^3}\,P_{\sigma}(q) \simeq
  {\cal P}_{\sigma}\,{\rm ln}\,\left(\frac{Q}{k_{IR}}\right)\ \,,
  \eeq
% where
%The ensemble average of long-wavelength
%fluctuations vanishes $\langle\vp_L\rangle=0$.
%This provides a convenient way to compute radiative corrections to
%inflationary observables.
Using the results above, %we find from
%equation (\ref{expan1}) that 
the curvature perturbation as measured
within a patch of volume $V_{\x_0}$ can be written as
  \bea
  \zeta(\x)_{|_{V_{\x_0}}}
 \simeq \sum_{n=1}^{\infty}\frac{N^{(n)}(\sigma_0+\delta\sigma_{L}(\x_0))}{n!}\,(\delta\sigma_s(\x))^n\ .\label{expand3}
 \eea
Converting this expression to Fourier space on scales $k > Q$ is
straightforward,  as $\delta\sigma_{L}(\x_0)$ acts as a constant under
this operation. However the LW fluctuations are operators
with non-vanishing ensemble, or full sky, 2-pt functions
(\ref{loge}).
%Indeed, an $n$-pt function evaluated in the small patch,
%$\langle\zeta(\k_1)...\zeta(\k_n)\rangle_{{\x_0}}$, corresponds to
%an operator containing fields $\delta\sigma_L(\x_0)$.
This has interesting consequences when considering full sky
expectation values of tree-level $n$-pt functions evaluated in
small patches $\langle\,\langle\zeta (\k_1)...\zeta (\k_n)\rangle_{{\x_0}}\,\rangle\,.$
We can expand the argument of $\langle \dots \rangle$ in terms of LW
modes $\delta \sigma_L$, and evaluate the ensemble averages: at this point
the connection with loops in inflation becomes apparent.
This operation is equivalent to computing radiative corrections to the
corresponding full sky $n$-pt functions in a leading-log
approximation. Ensemble averages of powers of $\delta \sigma_L$ lead
to log-enhanced contributions, controlled by formula (\ref{loge}).
By identifying the reference scale $Q$ with the scale $k$ at which
the measurement of full sky $n$-pt functions are performed, one
finds that the LW mode contributions {\it exactly} reproduce the
radiative corrections in the leading-log approximation.

%More specifically: both radiative corrections and LW contributions
%arise from the expansion of the curvature perturbation
%(\ref{expand3}) around the same classical background solution. The
%vertices appearing in the loop diagrams, and diagrams containing soft
%modes are by construction {\it identical} to the ones associated
%with LW corrections. Integrals over the loop momenta can be
%estimated using the so called pole approximation \cite{Boubekeur:2005fj}, which
%amounts to integrating each of the loop propagators from the IR
%cutoff $k_{\rm IR}$ up to the mode $k$ at which the corrections are
%computed. This produces $n$ logarithms of the form ${\rm
%ln}(\nicefrac{k}{k_{\rm IR}})$ where $n$ is the number of lines where the loop
%momentum appears. Exactly the same $n$ logarithms are reproduced by
%expanding the same correlator in long-wavelength modes provided that
%we set $Q\rightarrow k$. The same applies for integrals over soft
%modes. Indeed, since the radiative corrections consist of
%long-wavelength modes, and since we are performing a semiclassical
%leading logarithm expansion of the curvature perturbation, the exact
%correspondence is precisely what one expects to find.

This approach allows us to easily reproduce and extend our
discussion of radiative corrections to the equality (\ref{tfe}).
From the expansion (\ref{expand3}) we find the observables
$\bfnl^{\x_0}$ and $\btaunl^{\x_0}$, measuring respectively the
squeezed and collapsed limits of tree-level 3- and 4-pt
functions within a small patch $V_{\x_0}$, are given by
  \baq
 \bar{f}_{{\rm NL}}^{\x_0}&\equiv&\lim_{k_1\to0}\frac{5}{12}\,\frac{\bar{B}}{\bar{P}(k_1)
  \bar{P}(k_2)}=\frac{5}{6}\,\frac{N''(\sigma_0+\delta\sigma_L(\x_0))}{N'{}^2(\sigma_0+\delta \sigma_L(\x_0))}
  \ ,
  \label{pfnl}\\
  \label{ptnl}  \bar{\tau}_{{\rm
  NL}}^{\x_0}&\equiv&\lim_{{k_{12}}\to0}\frac{1}{4}\, \frac{\bar{T}}
  {\bar{P}(k_{12}) \bar{P}(k_1)
  \bar{P}(k_3)}=\frac{N''^2(\sigma_0+\delta \sigma_L(\x_0))}{N'{}^4(\sigma_0+\delta \sigma_L(\x_0))}\nonumber
  \ .
  \eaq
Taking ensemble averages
%of $\left(\bar{f}_{\rm NL}^{\x_0}\right)^2$
%and $ \bar{\tau}_{\rm NL}^{\x_0}$
we find,
% we then obtain the equality
  %\beq
  %\label{SYeq}
  $\langle\, \bar{\tau}_{{\rm
  NL}}^{\x_0}\,\rangle \, =\, \left(\nicefrac{6}{5}\right)^2\langle\,( \bar{f}_{{\rm
  NL}}^{\x_0})^2\,\rangle \neq \, \left(\nicefrac{6}{5}\right)^2\langle\, \bar{f}_{{\rm
  NL}}^{\x_0}\,\rangle^2 \ ,
 $
 % \eeq
which can also be expressed as
%$\langle B^2_{\x_0,\rm sq}
%\rangle \,=\, \langle T_{\x_0,\rm coll}\,P_{\x_0,\rm
%  soft}\rangle$, corresponding to
%    \baq
% \lim_{\substack{{k_1}{\to0}\\{{k_2}{\to k_3}\hspace{-4pt}}}} \langle
%  B^2_{{{\x_0}}}(k_1,k_2,k_3)\rangle &=& \lim_{\substack{{\k_1}{\to-\k_2}\\{k_1}{\to
%  k_3}\phantom{{\bf -}}\\{k_5}{\to 0}\hspace{10pt}}}  \langle
%  T_{{{\x_0}}}(k_1,k_2,k_3,k_4,k_{12},k_{13}) P_{{{\x_0}}}(k_5)
%  \rangle
%\,  . \label{b2tp}
%  \eaq
  \beq
  \langle \bar{B}^2_{\x_0}\rangle_{k_1\to 0} =
  \langle \bar{T}_{\x_0} \bar{P}_{\x_0} \rangle_{k_{12}\to 0,\,k_5\to 0}\ . \label{b2tp}
  \eeq

As discussed above, this relation  between tree-level quantities in
a small patch $V_{\x_0}$, equates {\it to all orders in radiative
corrections}, and to leading log accuracy, the corresponding
{\it full-sky quantities} evaluated  $k \sim
V_{\x_0}^{-\nicefrac{1}{3}}$. Expanding (\ref{b2tp}) to second order
in the LW modes $\delta \sigma_L$ exactly reproduces the first order
radiatively corrected equality between $(B^2)_{\rm rad} = (T P)_{\rm
rad}$, derived above using diagrammatic methods (\ref{newequa}). In
squaring $\bfnl^{\x_0}$ one obtains quadratic contributions in
$\delta\sigma_L$ both from the linear and quadratic terms in
equation (\ref{pfnl}). The former correspond to soft corrections,
and are the origin of the inhomogeneities of nonG discussed in
\cite{Byrnes:2011ri}, and the latter to loop corrections. Both of
them have to be included in order for the equality (\ref{newequa})
(or (\ref{b2tp})) to be satisfied. This is why the the loop
corrected quantities $\fnl^{\rm loop}$ and $\taunl^{\rm loop}$
(\ref{fnl_loop}), missing these soft contributions, in general fail
to satisfy the equality (\ref{tfe}), as seen in
(\ref{loop-violation}).

At one-loop order, it is instructive to rewrite
 (\ref{loop-violation})   as
  \baq
  \label{b2tp_variances}
&& \hskip -0.5cm  \langle\, B_{\x_0}\,\rangle^2_{k_1\to 0} =
\langle\, T_{\x_0} \,\rangle_{k_{12}\to 0}  \langle\,
P_{\x_0}\,\rangle_{k_5\to 0}
+\sqrt{\langle\sigma_T^2\sigma_P^2\rangle}-\sigma_B^2.
  \eaq
Here $\sigma_x^2=\langle x^2\rangle-\langle x\rangle^2$ denotes the
non-zero variance generated by the long-wavelength modes, giving a
statistical variation to quantities measured  on subhorizon patches.
 Writing the formula  in this form explicitly  shows that taking averages over full
 sky of single quantities, and then combining them together, does not lead to a simple
form of the equality. Additional pieces  proportional to $\bgnl$ (and $\bhnl$ at two loops) lead to a violation
 of the tree-level result (\ref{tfe}) when loops are included.
 Diagrammatically, as we have seen in the previous section,  these corrections
  can be traced back to soft
internal modes connecting  tree-level correlators. This can be
avoided by defining quantities that, once averaged over the full
sky, allow us to write an equality between the bispectrum and
trispectrum in a form that automatically handles radiative
 corrections at all orders, as given by Eq.~(\ref{b2tp}).

\smallskip

  \noindent
  {\bf Conclusions:}
   In this work we have investigated new contributions to local nonG inflationary observables in squeezed or collapsed configurations,
  associated with  soft momentum lines connecting different  $n$-pt functions.
   Our analysis is essential for investigating and  understanding
   consistency relations among inflationary observables,   that can
 be tested by the Planck satellite and provide model independent  information about the number of degrees of freedom contributing
 to the primordial curvature fluctuations.
   We showed that the new contributions we discussed  are essential
  for defining a  combination of power spectrum, bispectrum and trispectrum,  equation (\ref{b2tp}),  corresponding   to
   a new equality among observables
    preserved by radiative corrections in single-source inflationary scenarios. This provides the natural generalization  of the tree-level equality $ \bar{\tau}_{{\rm
NL}} \, =
\left(\nicefrac{6}{5} \bar{f}_{{\rm
NL}}\right)^2$, which is broken by loop corrections. We discussed our results adopting a convenient diagrammatic representation of inflationary $n$-pt functions  in terms of Feynman diagrams. We also made a connection between these results and  inhomogeneities of nonG observables, clarifying  the relation between loop corrections and inhomogeneous  nonG. In order to do this, we  employed  a particularly simple method based on splitting long from short wave-length modes with respect to a fiducial scale,  and exploited  the fact that  long wavelength  mode
   contributions to inflationary observables behave in  the same way as radiative  corrections.

   In summary, we have shown from various points of view that long wavelength, soft modes can provide physical contributions
   to inflationary observables.   
 In this work we focussed on scenarios characterized by large nonG, in which loop effects can provide sizeable corrections to the equality (\ref{tfe}).  Interestingly, our arguments can also be used to clarify puzzling results obtained in pure single field inflation, in which the level
  of nonG 
 is of order of slow roll parameters. In that case, it has been shown \cite{Seery:2008ax} that tree-level gravitational corrections to  $\tau_{NL}$ 
  give a contribution proportional to the tensor-to-scalar ratio $r=16 \epsilon$, while in the squeezed limit $\fnl$ is proportional to the tilt of the power spectrum $n_s-1=2 \eta-6 \epsilon$: hence, the equality  (\ref{tfe}) is violated because each side of that formula scales with a different power of the slow-roll parameters.  On the other hand, this does not happen for our new consistency relation (\ref{b2tp}).
  %, that is preserved even after including gravitational corrections.
    Indeed, a straightforward calculation \cite{preparation}
    shows  that, although gravitational waves do not contribute at tree level to the bispectrum $B$, they do to its square $B^2$.  The  contribution to $B^2$ is proportional to the tensor-to-scalar ratio, with the correct
 features 
   to  match with the trispectrum $T$ in the right hand side and preserve our consistency relation (\ref{b2tp}).     
    While in this paper we focussed on single-source scenarios,
    it is straightforward to generalize the method and our results to a
multiple-field case. The tree-level equality (\ref{tfe}) gets
replaced by the inequality $ \bar{\tau}_{{\rm
NL}} \, \geqslant
\left(\nicefrac{6}{5} \bar{f}_{{\rm
NL}}\right)^2$ \cite{Suyama:2007bg}. According to our previous
discussion, this translates into a new inequality between full-sky
observables:
  \beq\label{mfc}
  (B^2)^{\rm rad}_{k_1\to0} \leqslant ( P T)^{\rm
  rad}_{k_{12}\to0,\,k_5\to0}\ ,
  \eeq
which holds to all orders in radiative corrections, and to leading
logarithm precision.  This
inequality can be unambiguously used to discriminate between single
and multiple-source scenarios for generating primordial
perturbations at arbitrary orders in radiative corrections, even when loops are
included.
  The Suyama-Yamaguchi inequality and our inequality (\ref{mfc}) above are two observable relations probing
  different physics when loops or gravitational corrections are included, and capable of testing in different ways models
  which generate the curvature perturbation. We will explore this and other interesting issues
  elsewhere \cite{preparation}.

%{\bf GT I made the conclusions short, below is Sami's comment. I don't know if we need to talk about this here, or
%in a followup paper. I would prefer put in followup paper}

% {\bf SN:
%If the space allows, we should add a few words about the remaining
%degeneracy between single and multi source models. It would also be
%worth mentioning that the SY inequality and our inequality are both
%observables probing different physics. These insights along with the
%discussion of the variance below could be presented in the
%conclusions as they concern the practical use of our results.}

\smallskip

  \noindent
  {\bf Acknowledgments:}
GT is supported by an STFC Advanced Fellowship ST/H005498/1.
DW is supported by STFC grant ST/H002774/1.
SN and GT would like to thank CERN Theory Division for their warm hospitality.

\providecommand{\href}[2]{#2}\begingroup\raggedright

\end{document}